\documentclass[conference]{IEEEtran}
\IEEEoverridecommandlockouts

\usepackage{cite}
\usepackage{amsmath,amssymb,amsfonts}
\usepackage{algorithmic}
\usepackage{graphicx}
\usepackage{textcomp}
\usepackage[table]{xcolor}

\usepackage{booktabs}
\usepackage{multirow}
\usepackage{amsmath}
\usepackage{glossaries}
\usepackage[american]{circuitikz}
\newacronym{rms}{RMS}{root-mean-square}
\newacronym{emt}{EMT}{electromagnetic transient}
\newacronym{dae}{DAE}{differential-algebraic equation}
\newacronym{ode}{ODE}{ordinary differential equation}

\newcommand\blfootnote[1]{%
  \begingroup
  \renewcommand\thefootnote{}\footnote{#1}%
  \addtocounter{footnote}{-1}%
  \endgroup
}

\def\BibTeX{{\rm B\kern-.05em{\sc i\kern-.025em b}\kern-.08em
    T\kern-.1667em\lower.7ex\hbox{E}\kern-.125emX}}
\begin{document}

\title{Limitations of RMS-based Stability Assessment of Converter-based Grids: A Case Study}

\author{\IEEEauthorblockN{1\textsuperscript{st} Pol Jane-Soneira}
\IEEEauthorblockA{\textit{Corporate Research Center} \\
\textit{ABB}\\
Mannheim, Germany \\
pol.jane-soneira@de.abb.com}
\and
\IEEEauthorblockN{2\textsuperscript{nd} Ognjen Stanojev}
\IEEEauthorblockA{\textit{Corporate Research Center} \\
\textit{ABB}\\
Baden-Dättwil, Switzerland \\
ognjen.stanojev@ch.abb.com}
\and
\IEEEauthorblockN{3\textsuperscript{rd} Francisco Canales Perez}
\IEEEauthorblockA{\textit{Mining / Automation} \\
\textit{ABB}\\
Baden, Switzerland \\
francisco.canales-perez@ch.abb.com}
\and
\IEEEauthorblockN{4\textsuperscript{th} Qian Long}
\IEEEauthorblockA{\textit{Energy Industries} \\
\textit{ABB}\\
City, Norway \\
qian.long@no.abb.com}
\and
\IEEEauthorblockN{5\textsuperscript{th} Mario Schweizer}
\IEEEauthorblockA{\textit{Corporate Research Center} \\
\textit{ABB}\\
Baden-Dättwil, Switzerland \\
mario.schweizer@ch.abb.com}
\and
\IEEEauthorblockN{6\textsuperscript{th} Matthias Biskoping}
\IEEEauthorblockA{\textit{Corporate Research Center} \\
\textit{ABB}\\
Mannheim, Germany \\
matthias.biskoping@de.abb.com}
}

\maketitle

\blfootnote{This paper has been submitted to IEEE. Copyright may be transferred to IEEE.}

\begin{abstract}
The increasing penetration of power electronic converters in industrial grids introduces stability challenges at frequencies well above the electromechanical range of traditional power systems. Commercial software tools typically perform eigenvalue-based stability analysis using root-mean-square (RMS) models that assume a quasi-stationary network, thereby neglecting electromagnetic dynamics of assets --- which is standard practice in power system analysis. This paper presents a case study of an industrial grid where the conventional RMS-based eigenvalue analysis predicts stable operation, while a detailed electromagnetic transient (EMT) simulation reveals growing oscillations, indicating instability. To bridge this gap, we introduce an alternative modeling approach that formulates the network dynamics in the rotating $dq$ reference frame, retaining the differential equations of electromagnetic dynamics. The resulting model enables eigenvalue analysis that correctly identifies the unstable modes, consistent with the EMT simulation results. Our findings highlight a fundamental limitation of RMS-based stability assessment for converter-dominated grids and demonstrate that dynamic impedance models in the $dq$ frame provide a viable path toward accurate small-signal analysis encompassing high-frequency converter-network interactions.
\end{abstract}


\section{Introduction}

Industrial grids are undergoing a profound transformation driven by the increasing integration of power electronic converters, which are replacing traditional synchronous generators as the primary sources of power. This shift is motivated by environmental and economic aspects: on the one hand, the cost per kWh generated by renewable energy sources such as wind and solar has decreased significantly, and is 5-10 times lower than fossil fuel-based electricity~\cite{IRENA2024,kost2024levelized}. Furthermore, the cost of batteries has decreased considerably \cite{bajolle2022future}, enabling the use of energy storage systems to mitigate the variability of renewable generation. On the other hand, the carbon footprint of products is gaining importance in industrial sectors. For example, a carbon footprint declaration is mandatory for batteries larger than 2 kWh in the European Union~\cite{eu_regulation}. To be able to offer products with low carbon footprint, industries are increasingly striving to decarbonize their operations. Due to the aforementioned reasons, many industrial facilities are transitioning towards renewable generation-based grids, in which power electronic converters interface both the sources and, in many cases, also the loads.

However, the proliferation of converters also introduces new challenges in ensuring the stability of these grids, as they may interact with the passive network elements in ways that can give rise to high-frequency oscillations and other dynamic phenomena that have not been of concern in traditional power systems \cite{dorfler2023control,kong2021review,gross2019effect,buchhagen2015borwin1}. Furthermore, the lack of rotational inertia and damping leads to faster frequency dynamics and more frequent interactions between the units in the system. This calls for new stability analysis methods and metrics that allow the design of reliable and robust converter-based grids.

Commercial software tools like PowerFactory, PSS/E and Neplan allow for a largely automated stability analysis via eigenvalues and modal analysis \cite{powerfactory_user_manual, milano2017primal, neplan_web, kaberere2005comparative}. This is a well-known method that has been of great help in the past to fix stability issues in power systems \cite{Kundur1994, milano2017primal,kundur1990comprehensive}. Some of those commercial software packages have recently added new functionalities tailored to analyzing converter-based grids, e.g., PowerFactory with the impedance-based stability tool \cite{wurl2025multi}, \cite[Ch.~33]{powerfactory_user_manual}. 
However, these tools typically adopt the modeling assumption of a timescale separation between the transmission line dynamics and the generator or converter dynamics, which is standard in the analysis of traditional power systems~\cite{lara2023revisiting, Kundur1994, silva2015modal}. As we show later more in detail, this assumption may render the eigenvalue-based stability analysis unable to detect to high-frequency instabilities that can arise in converter-based grids.

In this paper, we present a case study of an industrial grid with an IEC Type~4A wind turbine, where the traditional eigenvalue-based stability analysis predicts stable operation, while a detailed electromagnetic transient simulation reveals growing high-frequency oscillations, indicating instability. Furthermore, we introduce an alternative modeling approach that retains the transmission line dynamics, enabling a more accurate stability assessment that captures the full dynamics of converter-based grids.
This example highlights the limitations of conventional stability analysis methods for power systems, adopted also by many commercial software tools, and underscores the need for alternative approaches that can capture the full dynamics of converter-based grids.

\section{Stability Analysis in Traditional Power Systems} \label{sec:stability_traditional}

In this section, we summarize how stability analysis is performed for traditional, synchronous generator based power systems. We begin with the system theoretic foundations of power system modeling and eigenvalue analysis, followed by a discussion of how this is implemented in commercial software tools. 

\subsection{Power System modeling and Eigenvalue Analysis}

\subsubsection{\Acrshort{rms} modeling}

In the \gls{rms} or phasor-based modeling framework, power systems are described by a set of \glspl{dae} of the form \cite{lara2023revisiting, sauer2014power}
\begin{subequations} \label{eq:dae}
    \begin{align}
        \dot{x} &= f(x, y), \label{eq:dae_diff} \\
        0 &= g(x, y), \label{eq:dae_alg}
    \end{align}
\end{subequations}
where $x \in \mathbb{R}^n$ denotes the vector of dynamic state variables (e.g., rotor angles, speeds, and controller states) and $y \in \mathbb{C}^m$ represents the vector of algebraic variables. The differential equations~\eqref{eq:dae_diff} capture the dynamics of individual components such as synchronous generators, their excitation systems, governors, and power electronic converters. The algebraic constraints~\eqref{eq:dae_alg} encode the network equations under the quasi-stationary assumption. Thus, voltages and currents are represented as complex phasors, and impedances are represented as complex quantities. Together with Kirchhoff's laws, this leads to algebraic equations that describe the physical network. 
This formulation assumes that electromagnetic transients within the network decay instantaneously relative to the electromechanical or controller time scales of interest, thereby eliminating the need to model transmission line dynamics~\cite{lara2023revisiting}. In addition, it is important to stress that the algebraic variables $y$ are not limited to network voltages and currents; they may also include other algebraic quantities such as flux linkages in generators and transformers, which are typically also treated as algebraic variables in an \gls{rms} framework \cite[Sec.~4-B]{lara2023revisiting}.


For small-signal stability analysis, the DAE system~\eqref{eq:dae} is linearized around an operating point $(x_0, y_0)$, yielding
\begin{align}
    \Delta \dot{x} &= A_{xx} \Delta x +  A_{xy} \Delta y, \\
    0 &=  A_{yx} \Delta x +  A_{yy} \Delta y,
\end{align}
where $A_{xx} = \frac{\partial f}{\partial x}\big|_{(x_0, y_0)}$, $A_{xy} = \frac{\partial f}{\partial y}\big|_{(x_0, y_0)}$, $A_{yx} = \frac{\partial g}{\partial x}\big|_{(x_0, y_0)}$, and $A_{yy} = \frac{\partial g}{\partial y}\big|_{(x_0, y_0)}$ are the Jacobian matrices evaluated at the operating point. 
If $A_{yy}$ is nonsingular\footnote{This is typically the case for normal operation, see \cite{grijalva2005necessary, sauer2014power}. If that is not the case, a generalized eigenvalue problem needs to be solved to assess small signal stability~\cite{milano2017primal}.}, the algebraic variables can be eliminated to obtain the reduced state matrix $A_{\mathrm{red}} = A_{xx} - A_{xy} A_{yy}^{-1} A_{yx}$, whose eigenvalues determine the small-signal stability of the system \cite{Kundur1994}. The system is stable if and only if all eigenvalues of $A_{\mathrm{red}}$ have strictly negative real parts.

\subsubsection{\Acrshort{emt} modeling}

In \gls{emt} modeling, the quasi-stationary network assumption is relaxed, and the transmission network is itself described by differential equations that capture fast electromagnetic phenomena such as travelling waves and resonances. Transmission lines, cables and other impedances are modeled through distributed or lumped parameter models, leading to differential equations of the form
\begin{equation}
    \dot{x}_{\mathrm{net}} = f_{\mathrm{net}}(x_{\mathrm{net}}, u_{\mathrm{net}}),
    \label{eq:emt_network}
\end{equation}
where $x_{\mathrm{net}}$ contains the inductor currents and capacitor voltages of the network elements, and $u_{\mathrm{net}}$ represents the terminal quantities interfacing with the dynamic components (i.e. voltages and currents). In contrast to the \gls{rms} framework, voltages and currents are represented as instantaneous time-domain sinusoidal quantities rather than complex phasors.

When the network dynamics~\eqref{eq:emt_network} are combined with the component models, the overall system takes the form of an \gls{ode}
\begin{equation}
    \dot{\tilde{x}} = \tilde{f}(\tilde{x}),
    \label{eq:emt_ode}
\end{equation}
where $\tilde{x} = [x^\top, x_{\mathrm{net}}^\top]^\top$ is the augmented state vector comprising both component and network states. In certain formulations, particularly when ideal transformers or other simplified elements are present, the system retains algebraic constraints and is expressed as a \gls{dae}. This \gls{ode} or \gls{dae} system characterizes the full electromagnetic and electromechanical dynamics of the power system, including high-frequency dynamics that are not described in the \gls{rms} framework.

Note, however, that a linearization of~\eqref{eq:emt_ode} around an equilibrium is not possible with the EMT system as described here. This is because there exists no constant equilibrium point for the EMT system, as the network states $x_{\mathrm{net}}$ are time-varying sinusoidal signals. Without an equilibrium point, the eigenvalues of the Jacobian matrix of the EMT system do not provide a meaningful stability analysis.

\subsection{Stability Analysis in Traditional Power Systems and Commercial Software Tools}

In traditional power systems, the dynamic behavior is predominantly governed by synchronous generators and their associated controllers (excitation systems, governors, and power system stabilizers). The stability phenomena of primary concern, such as inter-area oscillations, local plant modes, and loss of synchronism, are of electromechanical nature and occur at frequencies typically below a few hertz~\cite{Kundur1994}. High-frequency electromagnetic transients within the network were generally assumed to be well-damped and to decay rapidly, posing no threat to system stability. 
Furthermore, works such as~\cite{zaborszky1993error} demonstrated that \gls{rms} models provide an adequate representation of the system dynamics in networks dominated by synchronous generators, introducing negligible error compared to full electromagnetic models. 

Combined with the fundamental difficulty of performing small-signal analysis on \gls{emt} models---as discussed above, the absence of a constant equilibrium point prevents a direct linearization and eigenvalue computation---\gls{rms} models became the established standard for small-signal stability assessment in power systems~\cite{lara2023revisiting, vega2020analysis}.

Commercial software tools such as DIgSILENT PowerFactory, PSS/E, and Neplan follow this traditional methodology~\cite{silva2015modal, powerfactory_user_manual, psse_2013}. They perform eigenvalue and modal analysis based on \gls{rms} models, constructing the \gls{dae} system~\eqref{eq:dae} with a quasi-stationary network representation and computing the eigenvalues of the reduced state matrix $A_{\mathrm{red}}$. While this approach has proven highly effective for synchronous generator-dominated grids, it inherits the well-known modeling simplifications of the \gls{rms} framework: any dynamic phenomena arising from the interaction between converter controls and network electromagnetic dynamics at frequencies above the electromechanical are not captured.

\section{Example: Industrial Grid with IEC Type 4A Wind Turbine} \label{sec:case_study}

\begin{figure}[ht]
    \centering
    \includegraphics[width=\linewidth]{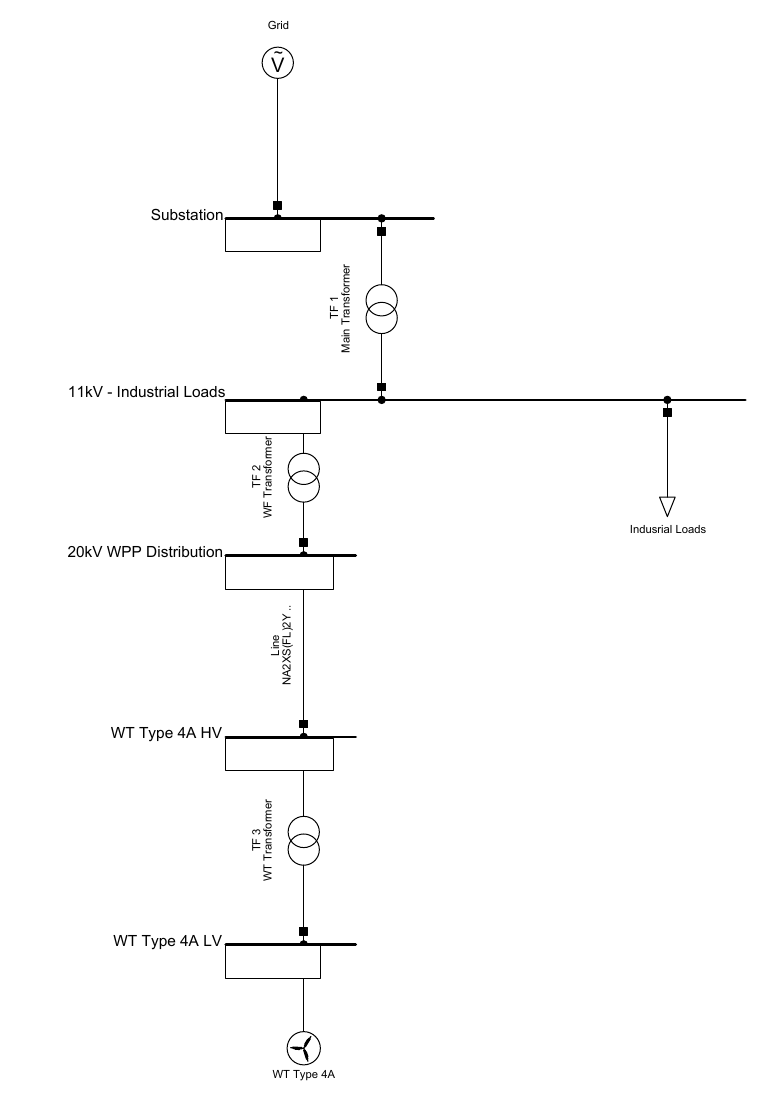}
    \caption{Grid configuration used in the case study.}
    \label{fig:grid}
\end{figure}

To illustrate the limitations of conventional stability analysis with \gls{rms} models, we consider an industrial grid fed by a IEC Type~4A wind turbine and an external grid, as depicted in Fig.~\ref{fig:grid}. Type 4A wind turbines are connected to the system via a full scale power electronic converter. The system comprises also three transformers (TF\,1--3), a transmission line, and a variable industrial load. The complete set of network parameters is given in Table~\ref{tab:network_params}. Details on the component models are provided in the next section.

This example is implemented in DIgSILENT PowerFactory 2025 SP4 using models from the standard library \cite{powerfactory_user_manual}. The system is then used to perform both the \gls{rms}-based eigenvalue analysis and the detailed \gls{emt} simulation. Note that we could have chosen any other commercial software package, and the results would be qualitatively the same, since the underlying \gls{rms} modeling assumptions are shared across all tools. 

\begin{table}[ht]
\centering
\caption{Network parameters.}
\label{tab:network_params}
\begin{tabular}{lll}
\toprule
\textbf{Description} & \textbf{Parameter} & \textbf{Value} \\
\midrule
\multicolumn{3}{c}{\textit{Transformers}} \\
\midrule
TF 1: Leakage & $R, L$ & 52.9 m$\Omega$, 1.675 mH \\
TF 1: Magnetizing & $R, L$ & 2.645 k$\Omega$, 4.86 H  \\
TF 1: turn ratio & $n$ & 33/11 \\
\midrule
TF 2: Leakage & $R, L$ & 55.56 m$\Omega$, 2.1147 mH  \\
TF 2: Magnetizing & $R, L$ & 4 k$\Omega$, 4.5 H  \\
TF 2: turn ratio & $n$ & 11/20 \\
\midrule
TF 3: Leakage & $R, L$ & 1.68 m$\Omega$, 35.98 mH \\
TF 3: Magnetizing & $L$ & 0.6 H  \\
TF 3: turn ratio & $n$ & 20/0.7 \\
\midrule
\multicolumn{3}{c}{\textit{Power Line}} \\
\midrule
series impedance & $R, L$ & 0.1289 $\Omega$, 0.384 mH  \\
parallel capacitance & $C$ & 0.228 $\mu$F  \\
\midrule
\multicolumn{3}{c}{\textit{WT Type 4A}} \\
\midrule
Inner current control & $k_{p}, k_{i}$ & 1e-3, 0.1  \\
PLL & $T_{pll}, u_{pll,1}, u_{pll,2}$ & 0.05, 0.05, 0.01  \\
Filter & $L$ & 0.0758 mH\\
\midrule
\multicolumn{3}{c}{\textit{Grid Connection}} \\
\midrule
Voltage amplitude & $v$ & 33 kV  \\
voltage angle & $\delta$ & 0 \\
\midrule
\multicolumn{3}{c}{\textit{Industrial Loads}} \\
\midrule
General Load & $P, Q$ & 0...2 MW, 0 MW  \\
\bottomrule
\end{tabular}
\end{table}

\subsection{Network and Component Models}

The grid consists of an industrial bus which feeds industrial loads, a wind turbine and a grid connection. The wind turbine of IEC Type~4A is connected through TF\,3 to a wind power plant distribution system at 20 kV, which includes a transmission line, and finally a transformer TF\,2 which connects to the industrial bus (see Fig.~\ref{fig:grid}). The industrial bus is also connected to an external grid via transformer TF\,1. The following subsections describe the models employed for each component more in detail.

\subsubsection{Transformers}

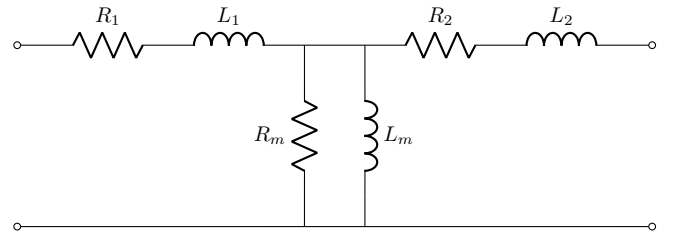
\begin{figure}[ht]
    \centering
    \begin{circuitikz}[scale=0.8, transform shape]
        \draw (0,0) to[short, o-] (0.5,0)
            to[R, l=$R_1$] (2.5,0)
            to[L, l=$L_1$] (4.5,0) -- (6,0);
        \draw (6,0) to[R, l=$R_2$] (8,0)
            to[L, l=$L_2$] (10,0)
            to[short, -o] (10.5,0);
        \draw (4.75,0) -- (4.75,-0.5)
            to[R, l_=$R_m$] (4.75,-2.5) -- (4.75,-3);
        \draw (5.75,0) -- (5.75,-0.5)
            to[L, l=$L_m$] (5.75,-2.5);
        \draw (5.75,-2.5) -- (5.75,-3);
        \draw (0,-3) to[short, o-] (4.5,-3)
            to[short, -o] (10.5,-3);
    \end{circuitikz}
    \caption{Transformer equivalent circuit.}
    \label{fig:transformer_circuit}
\end{figure}

The transformer equivalent circuit is shown in Fig.~\ref{fig:transformer_circuit}. Each transformer is modeled with a series branch representing the leakage impedance, and a parallel branch representing the magnetizing impedance. The leakage impedance consists of resistors $R_1$, $R_2$ and inductors $L_1$, $L_2$ in series, while the parallel branch consists of a resistor $R_m$ and an inductor $L_m$ in parallel. The turn ratio $n$ is also included to capture the voltage transformation between the primary and secondary sides (not displayed in Fig.~\ref{fig:transformer_circuit}). The specific parameters for each transformer are given in Table~\ref{tab:network_params}.

Note that within the \gls{rms} framework, the transformer is modeled using only complex-valued algebraic equations, since it is composed only of impedances. Hence, the transformer is modeled with equations~\eqref{eq:dae_alg}.

\subsubsection{Transmission Line}

\begin{figure}[ht]
    \centering
    \begin{circuitikz}[scale=0.8, transform shape]
        \draw (-0.5,0) to[short, o-] (1,0)
            to[R, l=$R_1$] (3,0)
            to[L, l=$L_1$] (5,0) to[short, -o] (6.5,0);
        \draw (5.5,0) -- (5.5,-0.5)
            to[C, l_=$\frac{C}{2}$] (5.5,-2.5) -- (5.5,-3);
        \draw (0.5,0) -- (0.5,-0.5)
            to[C, l_=$\frac{C}{2}$] (0.5,-2.5) -- (0.5,-3);
        \draw (-0.5,-3) to[short, o-] (4.5,-3)
            to[short, -o] (6.5,-3);
    \end{circuitikz}
    \caption{Transmission line equivalent circuit.}
    \label{fig:line_circuit}
\end{figure}
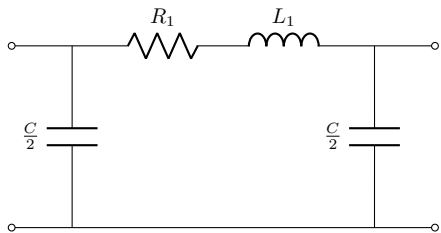

The transmission line connecting the wind turbine distribution system to the industrial bus is represented by a lumped-parameter $\pi$-equivalent circuit, as shown in Fig.~\ref{fig:line_circuit}. The series branch consists of a resistance $R$ and an inductance $L$ that account for the conductor losses and the magnetic field energy storage, respectively. The shunt capacitances $C/2$ at each end of the line represent the electric field energy stored between the conductors and ground. Similar to the transformer, the transmission line is modeled using only algebraic equations within the \gls{rms} framework, whereas the full differential equations governing the line capacitances and inductance are used in the \gls{emt} simulation.

\subsubsection{IEC Type 4A Wind Turbine}

For the wind turbine, the standard IEC Type~4A model from the PowerFactory library is used. More details of the model are available in \cite{IEC_WT_models,VillenaRuiz2022}. In the \gls{rms} model, the current control is simplified and represented by a first-order delay. The PLL is modeled as a second-order system with specified time constants. The converter output filter is represented by its reactance. In the \gls{emt} simulation, the current control implements PI controllers and coordinate rotations without simplifications, and the filter dynamics are captured through its respective differential equations.

\subsubsection{Industrial Load}

For the industrial load, a simple resistive load is used, computed such that it matches the given power consumption (see Table~\ref{tab:network_params}). This modeling may not be accurate enough for many use cases. In this paper, however, the focus is to showcase the converter-to-grid interactions which may occur in converter-based grids. Those interactions are already present due to the wind turbine, and thus a resistive load is used for simplicity. In general, resistive loads do not present the worst case from stability perspective as they provide additional damping that might not be present with equivalent converter-interfaced loads.

\subsection{Eigenvalue-Based Stability Analysis}

The small-signal stability of the system is assessed using the built-in modal analysis tool of DIgSILENT PowerFactory 2025 SP4. Following the \gls{rms}-based approach described in Section~II, which is standard in power systems, PowerFactory constructs the DAE system from the component models and the quasi-stationary network equations, linearizes around the operating point, and computes the system eigenvalues.

\begin{figure}[ht]
    \centering
    \includegraphics[width=1\linewidth]{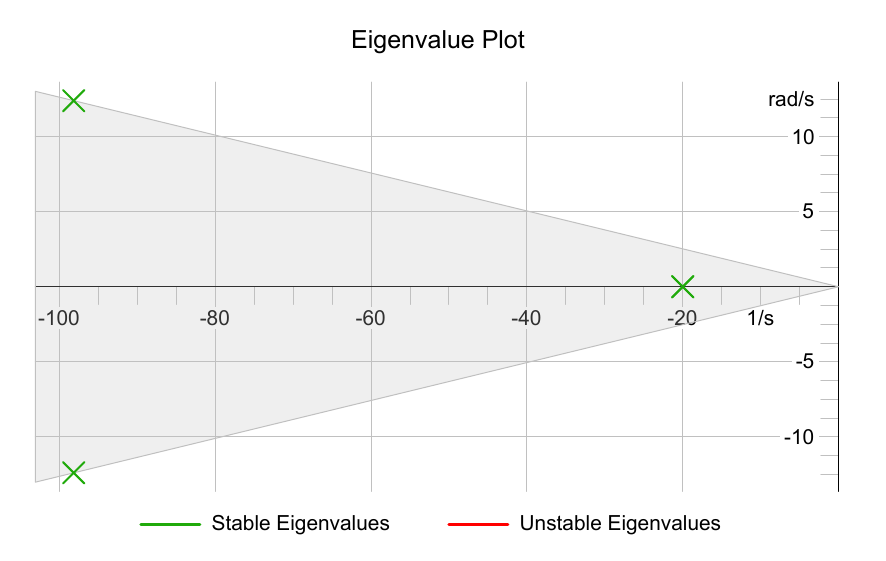}
    \caption{Eigenvalues computed using DIgSILENT PowerFactory for zero load. All eigenvalues lie in the left half-plane, indicating stability.}
    \label{fig:eigenvalues_rms_simplified_wt_stable}
\end{figure}

Fig.~\ref{fig:eigenvalues_rms_simplified_wt_stable} shows the resulting eigenvalue map for the no-load operating condition ($P = 0$). There are four eigenvalues, two corresponding to the simplified PI current control and two to the PLL, respectively. All eigenvalues are located in the open left half-plane, indicating that the system is small-signal stable according to this analysis. No oscillatory modes with positive real parts are detected, and the dominant modes exhibit adequate damping ratios. The eigenvalues do not change for any value of the industrial load $P$ as in Table~\ref{tab:network_params}. Based on this result alone, one would conclude that the system operates in a stable manner. 

\subsection{EMT Simulation}

To validate the stability prediction obtained from the eigenvalue analysis, a detailed EMT simulation of the same system is performed. In contrast to the \gls{rms}-based analysis, the EMT simulation captures the full electromagnetic dynamics of the network, including high-frequency interactions between the converter control and the passive network elements.

\begin{figure}
    \centering
    \includegraphics[width=1\linewidth]{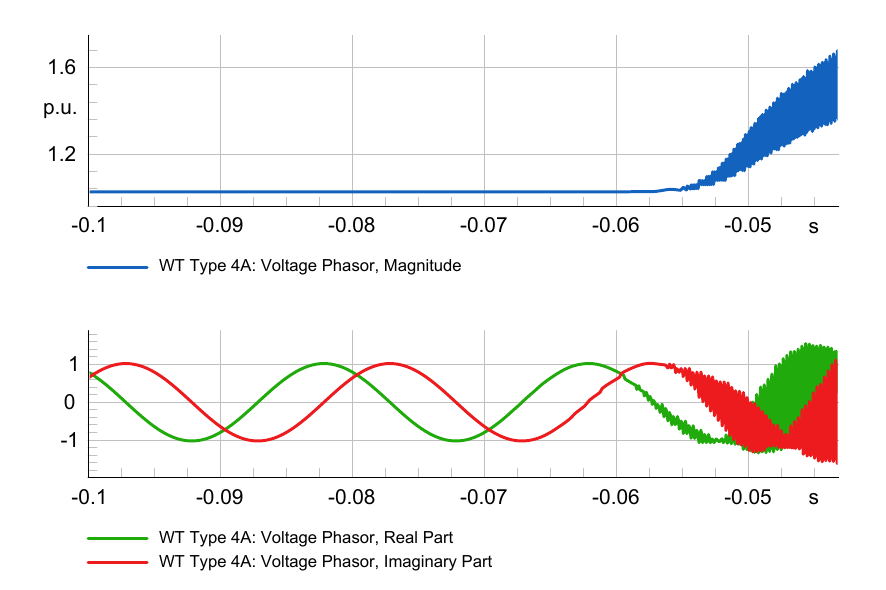}
    \caption{Voltage amplitude (above) and $\alpha\beta$-components (below) over time at the wind turbine transformer under zero load. The simulation starts at $t = -0.1$s by convention.}
    \label{fig:emt_voltage_alpha_beta_unstable}
\end{figure}

\begin{figure}
    \centering
    \includegraphics[width=1\linewidth]{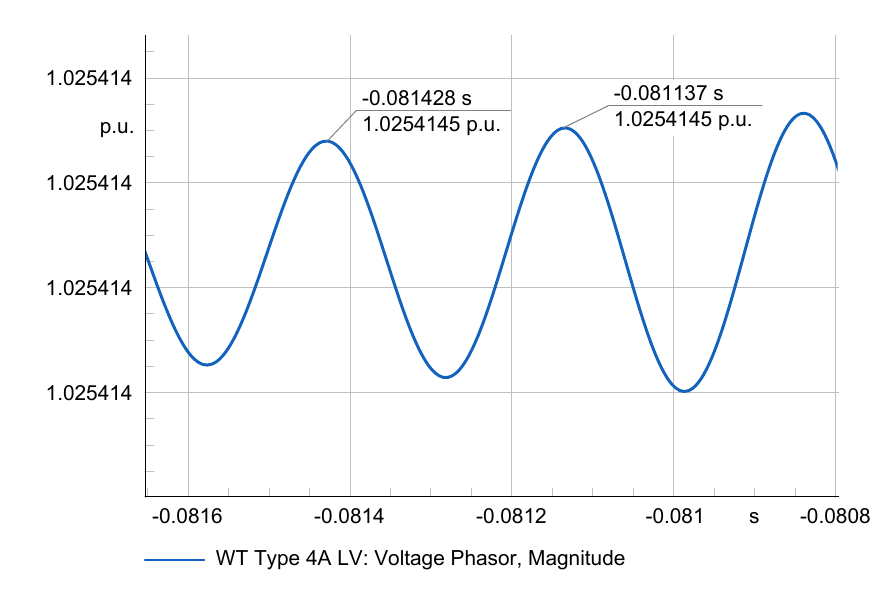}
    \caption{Zoomed view of the voltage amplitude over time, revealing oscillations at approximately 3.43 kHz.}
    \label{fig:emt_zoom_frequency_unstable}
\end{figure}

Fig.~\ref{fig:emt_voltage_alpha_beta_unstable} presents the simulated voltage at the wind turbine transformer in both amplitude and $\alpha\beta$-components. Despite the eigenvalue analysis predicting stable operation, the EMT simulation reveals growing oscillations in the voltage waveform. The oscillation amplitude increases over time, clearly indicating an unstable operating condition. A closer inspection of the oscillation frequency, shown in Fig.~\ref{fig:emt_zoom_frequency_unstable}, reveals that the instability occurs at approximately  
\begin{equation}
    f_{\mathrm{osc}} = \frac{1}{0.081428 \mathrm{s} - 0.081137 \mathrm{s}} \approx 3.436 \mathrm{kHz},
\end{equation}
a frequency well above the electromechanical range captured with \gls{rms} models.

\subsection{Discussion}

Significant divergence is observed between the eigenvalue-based stability assessment and the EMT simulation outcomes: the former predicts a stable system, while the latter demonstrates clear instability. This contradiction is a direct consequence of the underlying modeling assumptions inherent to the \gls{rms} framework.

As discussed in Section~\ref{sec:stability_traditional}, \gls{rms}-based stability analysis treats the network as quasi-stationary, representing it through purely algebraic equations. This assumption is well-justified in traditional power systems dominated by synchronous generators, where the relevant dynamics, e.g., electromechanical oscillations, voltage regulation, and governor response, occur at frequencies well below the electromagnetic bandwidth of the network. Under these conditions, network transients decay rapidly relative to the phenomena of interest, and the algebraic network representation introduces negligible error.

However, in converter-based grids, the situation is fundamentally different. Power electronic converters operate with control bandwidths that can extend into the kilohertz range, where they interact with the electromagnetic dynamics of transformers, cables, and transmission lines. These interactions give rise to oscillatory modes at frequencies that are invisible to the \gls{rms} framework, precisely because the network dynamics have been eliminated by assumption. The unstable mode at 3.43\,kHz observed in the EMT simulation is a manifestation of such a converter-network interaction, and no \gls{rms}-based eigenvalue analysis can detect it.

It is important to note that the limitations observed here are not specific to any particular software tool, but are shared by all stability analysis approaches that rely on the quasi-stationary network\footnote{We want to stress that the quasi-stationary assumption is not limited to the transmission lines, but also impedances of generators, transformers, and other elements in the network.} assumption. For grids with a high penetration of power electronic converters, alternative modeling frameworks that retain the network dynamics are required to ensure a reliable stability assessment.

\section{Alternative Modeling Approach for Stability Analysis} \label{sec:A-RMS}

The previous section demonstrated that the quasi-stationary network assumption renders \gls{rms}-based eigenvalue analysis unable to detect to high-frequency instabilities in converter-dominated grids. At the same time, a direct linearization of the \gls{emt} system is not feasible because the sinusoidal steady-state signals do not constitute a constant equilibrium point. To overcome both limitations, this section introduces an alternative modeling approach that formulates the network dynamics in the synchronously rotating $dq$ reference frame, as proposed in~\cite{pogaku2007modeling, katiraei2007small, iyer2010generalized}. In this frame, balanced three-phase sinusoidal quantities at fundamental frequency map to constant values in steady state, thereby providing a well-defined equilibrium suitable for linearization. Simultaneously, the differential equations governing inductances and capacitances are retained, so that the electromagnetic dynamics of the network are captured.

\subsection{Dynamic Impedances in the $dq$ Frame}

In the $dq$ reference frame rotating at the synchronous angular frequency $\omega_0$, the voltage--current relationship of an inductance $L \in \mathbb{R}_{>0}$ with resistance $R \in \mathbb{R}_{\geq 0}$ is given by the coupled differential equations
\begin{subequations} \label{eq:inductor_dq}
    \begin{align}
        L \frac{d i_d}{dt} &= v_d - R\, i_d + \omega_0 L\, i_q, \\
        L \frac{d i_q}{dt} &= v_q - R\, i_q - \omega_0 L\, i_d,
    \end{align}
\end{subequations}
where $v_d$, $v_q$ are the $d$- and $q$-axis voltages across the element, and $i_d$, $i_q$ are the corresponding currents. The cross-coupling terms $\pm \omega_0 L\, i_{q,d}$ arise from the coordinate transformation.

Similarly, a capacitance $C \in \mathbb{R}_{>0}$ is described by
\begin{subequations} \label{eq:capacitor_dq}
    \begin{align}
        C \frac{d v_d}{dt} &= i_d + \omega_0 C\, v_q, \\
        C \frac{d v_q}{dt} &= i_q - \omega_0 C\, v_d,
    \end{align}
\end{subequations}
where the capacitor voltages $v_d$, $v_q$ are now dynamic states and the injected currents $i_d$, $i_q$ serve as inputs.

In steady state at frequency $\omega_0$, all derivatives in~\eqref{eq:inductor_dq} and~\eqref{eq:capacitor_dq} vanish, yielding the expected constant $dq$ quantities. The key advantage is that perturbations around this equilibrium, including those at frequencies far from the fundamental, are captured by the differential equations, enabling eigenvalue-based small-signal analysis that includes electromagnetic dynamics.

This formulation is applied systematically to all passive network elements in the case study: the leakage and magnetizing impedances of transformers TF\,1, TF\,2, and TF\,3, the transmission line impedances, and the converter output filter. All are modeled by its $dq$-frame differential equations~\eqref{eq:inductor_dq} and~\eqref{eq:capacitor_dq}. The resulting system, comprising both the converter control states and the network electromagnetic states in the $dq$ frame, is a system that can be linearized around the steady-state operating point to obtain a state matrix whose eigenvalues reflect the correct dynamics in the kHz range.

\subsection{Eigenvalue Analysis with Dynamic Impedances}

\begin{figure}
    \centering
    \includegraphics[width=1\linewidth]{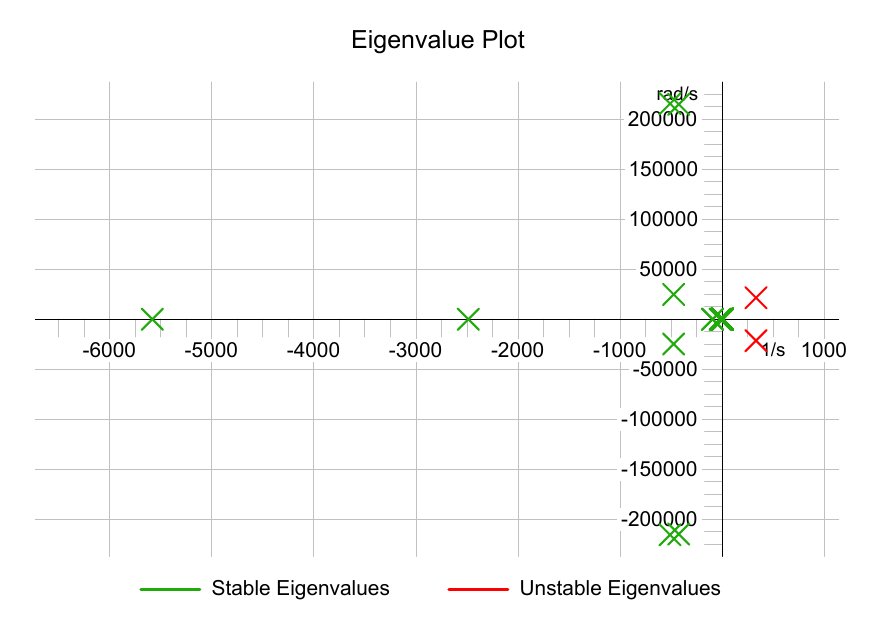}
    \caption{Eigenvalues computed using custom models with zero load. Unstable eigenvalue at $\lambda = 330\pm 21457.9j$}
    \label{fig:eigenvalues_a_rms_simplified_wt_unstable}
\end{figure}

The linearized system obtained using the $dq$-frame models is used to compute the eigenvalues of the full state matrix, which now includes both the converter control states and the electromagnetic states of the network. Fig.~\ref{fig:eigenvalues_a_rms_simplified_wt_unstable} shows the resulting eigenvalue map for the same zero-load operating condition considered in Section~\ref{sec:case_study}. In contrast to the \gls{rms}-based analysis, which predicted stability, the $dq$-frame models reveal a pair of complex conjugate eigenvalues with a positive real part at $\lambda = 330 \pm 21457.9j$. The imaginary part corresponds to a frequency of $f = 21457.9 / (2\pi) \approx 3.41$\,kHz, which is in line with the oscillation frequency of approximately 3.43\,kHz observed in the EMT simulation. This confirms that the $dq$-frame dynamic impedance model correctly captures the unstable converter-network interaction that the \gls{rms} framework fails to detect.

\begin{figure}[h]
    \centering
    \includegraphics[width=\linewidth, page=2]{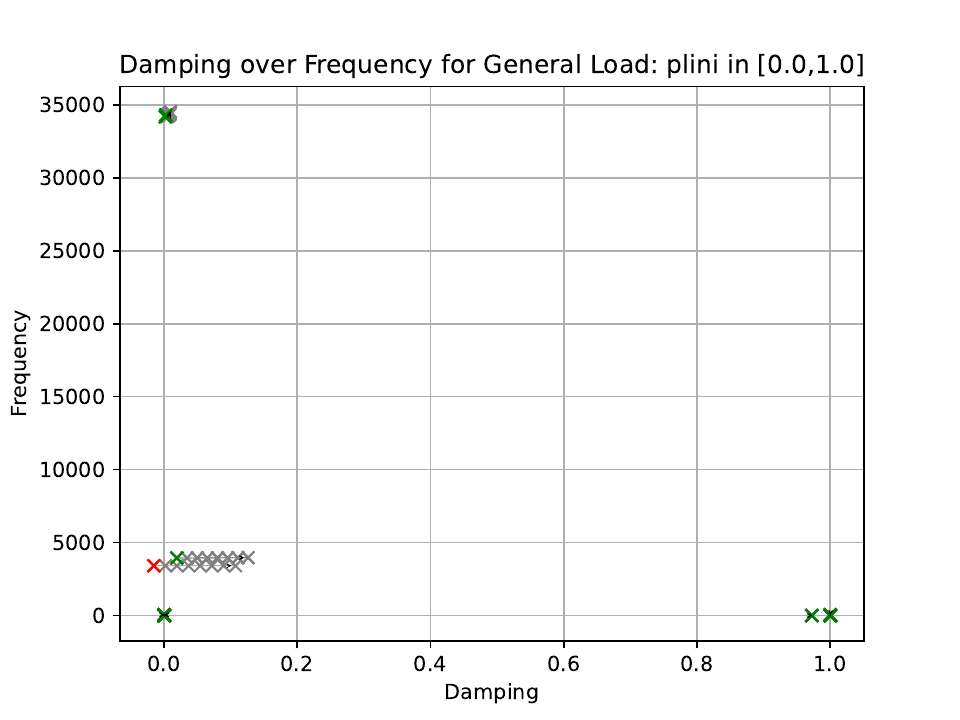}
    \caption{Parameter sweep of load power from 0 to 1 MW.}
    \label{fig:param_sweep_load}
\end{figure}

To further investigate the stability properties of the grid, a parameter sweep of the load power is performed from 0 to 1\,MW, and the eigenvalue trajectories are shown in Fig.~\ref{fig:param_sweep_load}. As the load increases, the critical eigenvalue pair moves towards the positive damping region. This behavior is consistent with the physical intuition that an additional resistive load provides damping to the resonant interaction between the converter and the network impedances. Note, however, that in a typical industrial grid, the load may not be resistive, but interfaced via a converter, which may not provide the same damping effect.

Overall, the $dq$-frame dynamic impedance model enables a comprehensive eigenvalue-based stability analysis. Thus, it allows practitioners to identify and mitigate potential instabilities relying on well known small-signal analysis techniques (e.g. participation factors). This may considerably reduce the effort for solving stability issues detected during EMT simulations, without the need for blindly trying out parametrizations in the computationally intensive EMT simulations. Recently, research groups in academia have also identified this issue and are starting to develop open-source toolboxes tackling this issue~\cite{arevalosoler2025stamp,serrano2022stability}. However, to the best of the authors' knowledge, commercial software tools do not yet offer this type of small-signal analysis with dynamic impedances in the $dq$-frame.

\section{Conclusion}

This paper presented a case study of an industrial grid with an IEC Type~4A wind turbine, in which we showed that the quasi-stationary network assumption --- which is standard practice in power system stability analysis --- underlying all commercial \gls{rms}-based tools means that their modal analysis, by design, does not capture high-frequency converter-network interactions. Specifically, the conventional eigenvalue analysis predicted stable operation, while the \gls{emt} simulation revealed a growing oscillation.

To address this gap, we introduced a modeling approach based on dynamic impedances formulated in the synchronously rotating $dq$ reference frame. By retaining the differential equations of all passive network elements, this framework provides a well-defined equilibrium suitable for linearization while capturing electromagnetic dynamics up to the kilohertz range. The resulting eigenvalue analysis correctly identified the unstable mode at 3.41\,kHz, consistent with the \gls{emt} simulation.

These findings underscore that, as industrial grids transition towards high converter penetration, stability assessment methodologies must evolve beyond the quasi-stationary assumptions inherited from synchronous generator-dominated systems. The $dq$-frame dynamic impedance approach offers a viable path for accurate small-signal analysis that leverages established eigenvalue techniques while encompassing a larger frequency range of converter-network interactions.

\bibliographystyle{IEEEtran}
\bibliography{IEEEexample}

@misc{arevalosoler2025stamp,
  title         = {A Matlab-based Toolbox for Automatic EMT Modeling and Small-Signal Stability Analysis of Modern Power Systems},
  author        = {Josep Arevalo-Soler and Dionysios Moutevelis and Elia Mateu-Barriendos and Onur Alican and Carlos Collados-Rodriguez and Marc Cheah-Mañe and Eduardo Prieto-Araujo and Oriol Gomis-Bellmunt},
  year          = {2025},
  eprint        = {2506.22201},
  archiveprefix = {arXiv},
  primaryclass  = {eess.SY},
  url           = {https://arxiv.org/abs/2506.22201}
}

@article{VillenaRuiz2022,
  title   = {Extensive model validation for generic IEC 61400-27-1 wind turbine models},
  journal = {International Journal of Electrical Power \& Energy Systems},
  volume  = {134},
  pages   = {107331},
  year    = {2022},
  issn    = {0142-0615},
  doi     = {https://doi.org/10.1016/j.ijepes.2021.107331},
  author  = {R. Villena-Ruiz and A. Honrubia-Escribano and F. Jiménez-Buendía and J.L. Sosa-Avendaño and S. Frahm and P. Gartmann and J. Fortmann and P.E. Sørensen and E. Gómez-Lázaro}
}

@article{IEC_WT_models,
  author    = {{International Electrotechnical Commission}},
  publisher = {},
  title     = {{IEC 61400-27-1: Wind energy generation systems - Part 27-1: Electrical simulation models - Generic models}},
  year      = {2020},
  isbn      = {978-2-8322-8505-3},
  journal   = {IEC Standard}
}

@article{dorfler2023control,
  title     = {Control of low-inertia power systems},
  author    = {D{\"o}rfler, Florian and Gro{\ss}, Dominic},
  journal   = {Annual Review of Control, Robotics, and Autonomous Systems},
  volume    = {6},
  number    = {1},
  pages     = {415--445},
  year      = {2023},
  publisher = {Annual Reviews}
}

@article{gross2019effect,
  title     = {The effect of transmission-line dynamics on grid-forming dispatchable virtual oscillator control},
  author    = {Gro{\ss}, Dominic and Colombino, Marcello and Brouillon, Jean-S{\'e}bastien and D{\"o}rfler, Florian},
  journal   = {IEEE Transactions on Control of Network Systems},
  volume    = {6},
  number    = {3},
  pages     = {1148--1160},
  year      = {2019},
  publisher = {IEEE}
}

@article{kong2021review,
  title     = {Review of small-signal converter-driven stability issues in power systems},
  author    = {Kong, Le and Xue, Yaosuo and Qiao, Liang and Wang, Fei},
  journal   = {IEEE Open Access Journal of Power and Energy},
  volume    = {9},
  pages     = {29--41},
  year      = {2021},
  publisher = {IEEE}
}

@book{powerfactory_user_manual,
  author    = {PowerFactory},
  editor    = {},
  publisher = {DIgSILENT Gmbh},
  title     = {User Manual},
  year      = {2025}
}

@inproceedings{buchhagen2015borwin1,
  title        = {BorWin1-First Experiences with harmonic interactions in converter dominated grids},
  author       = {Buchhagen, Christoph and Rauscher, Christian and Menze, Andreas and Jung, Jochen},
  booktitle    = {International ETG Congress 2015; Die Energiewende-Blueprints for the new energy age},
  pages        = {1--7},
  year         = {2015},
  organization = {VDE}
}

@inproceedings{wurl2025multi,
  title        = {Multi-connection assessment for impedance-based stability analysis through power system impedance determination in the frequency domain},
  author       = {W{\"u}rl, Thomas and Weise, Bernd},
  booktitle    = {24th Wind \& Solar Integration Workshop (WISO 2025)},
  volume       = {2025},
  pages        = {644--651},
  year         = {2025},
  organization = {IET}
}

@misc{IRENA2024,
  author = {{International Renewable Energy Agency}},
  isbn   = {978-92-9260-669-5},
  title  = {Renewable Power Generation Costs in 2024},
  year   = {2024}
}

@article{kost2024levelized,
  title   = {Levelized Cost of Electricity Renewable Energy Technologies},
  author  = {Kost, Christoph and Müller, Paul and Sepúlveda Schweiger, Jael and Fluri, Verena and Thomsen, Jessica},
  year    = {2024},
  journal = {Fraunhofer ISE}
}

@book{Kundur1994,
  author    = {P. Kundur},
  isbn      = {007035958X},
  publisher = {McGraw-Hill},
  title     = {Power System Stability and Control},
  year      = {1994}
}

@article{eu_regulation,
  author  = {European Parliament and Commission},
  title   = {REGULATION (EU) 2023/1542},
  journal = {Official Journal of the European Union},
  year    = 2023
}

@article{bajolle2022future,
  title     = {The future of lithium-ion batteries: Exploring expert conceptions, market trends, and price scenarios},
  author    = {Bajolle, Hadrien and Lagadic, Marion and Louvet, Nicolas},
  journal   = {Energy Research \& Social Science},
  volume    = {93},
  pages     = {102850},
  year      = {2022},
  publisher = {Elsevier}
}

@article{milano2017primal,
  title     = {Primal and dual generalized eigenvalue problems for power systems small-signal stability analysis},
  author    = {Milano, Federico and Dassios, Ioannis},
  journal   = {IEEE Transactions on Power Systems},
  volume    = {32},
  number    = {6},
  pages     = {4626--4635},
  year      = {2017},
  publisher = {IEEE}
}

@article{serrano2022stability,
  title     = {Stability tool for electric power systems with a high penetration of electronic power converters},
  author    = {Serrano-Jim{\'e}nez, Daniel and Unamuno, Eneko and Gil-de-Muro, A and Aragon, DA and Ceballos, S and Barrena, JA},
  journal   = {Electric Power Systems Research},
  volume    = {210},
  pages     = {108115},
  year      = {2022},
  publisher = {Elsevier}
}

@article{lara2023revisiting,
  title     = {Revisiting power systems time-domain simulation methods and models},
  author    = {Lara, Jose Daniel and Henriquez-Auba, Rodrigo and Ramasubramanian, Deepak and Dhople, Sairaj and Callaway, Duncan S and Sanders, Seth},
  journal   = {IEEE Transactions on Power systems},
  volume    = {39},
  number    = {2},
  pages     = {2421--2437},
  year      = {2023},
  publisher = {IEEE}
}

@article{kundur1990comprehensive,
  title     = {A comprehensive computer program package for small signal stability analysis of power systems},
  author    = {Kundur, P and Rogers, GJ and Wong, DY and Wang, L and Lauby, MG},
  journal   = {IEEE Transactions on Power Systems},
  volume    = {5},
  number    = {4},
  pages     = {1076--1083},
  year      = {1990},
  publisher = {IEEE}
}

@inproceedings{zaborszky1993error,
  title     = {Error estimation and limitation of the quasi stationary phasor dynamics},
  author    = {Zaborszky, J and Schattler, H and Venkatasubramaninan, V},
  booktitle = {Proc. Power Syst. Comput. Conf},
  pages     = {721--729},
  year      = {1993}
}

@article{vega2020analysis,
  title     = {Analysis and application of quasi-static and dynamic phasor calculus for stability assessment of integrated power electric and electronic systems},
  author    = {Vega-Herrera, Jorge and Rahmann, Claudia and Valencia, Felipe and Strunz, Kai},
  journal   = {IEEE Transactions on Power Systems},
  volume    = {36},
  number    = {3},
  pages     = {1750--1760},
  year      = {2020},
  publisher = {IEEE}
}

@misc{neplan_web,
  url    = {https://neplan.ch/description/small-signal-stability-2/},
  author = {{Neplan AG}},
  title  = {Small Signal Stability}
}

@misc{psse_2013,
  title     = {PSSE 33.4 Model Library},
  note      = {},
  year      = {2013},
  publisher = {Siemens Industry, Inc.},
  address   = {Schenectady, NY}
}

@inproceedings{kaberere2005comparative,
  title     = {Comparative analysis and numerical validation of industrial-grade power system simulation tools: Application to small-signal stability},
  author    = {Kaberere, KK and Folly, KA and Ntombela, M and Petroianu, AI and others},
  booktitle = {Proceedings of the 15th Power Systems Computation Conference},
  pages     = {32--3},
  year      = {2005}
}

@phdthesis{silva2015modal,
  title  = {Modal analysis applied to the stability study of hydroelectric systems with modular structures},
  author = {Silva, Pedro Camilo Oliveira},
  year   = {2015},
  school = {EPFL}
}

@incollection{sauer2014power,
  title     = {Power system dynamic equilibrium, power flow, and steady-state stability},
  author    = {Sauer, Peter W and Pai, MA},
  booktitle = {Real-Time Stability in Power Systems: Techniques for Early Detection of the Risk of Blackout},
  pages     = {1--26},
  year      = {2014},
  publisher = {Springer}
}

@article{grijalva2005necessary,
  title     = {A necessary condition for power flow Jacobian singularity based on branch complex flows},
  author    = {Grijalva, Santiago and Sauer, Peter W},
  journal   = {IEEE Transactions on Circuits and Systems I: Regular Papers},
  volume    = {52},
  number    = {7},
  pages     = {1406--1413},
  year      = {2005},
  publisher = {IEEE}
}

@article{pogaku2007modeling,
  title     = {Modeling, analysis and testing of autonomous operation of an inverter-based microgrid},
  author    = {Pogaku, Nagaraju and Prodanovic, Milan and Green, Timothy C},
  journal   = {IEEE Transactions on power electronics},
  volume    = {22},
  number    = {2},
  pages     = {613--625},
  year      = {2007},
  publisher = {IEEE}
}

@article{katiraei2007small,
  title     = {Small-signal dynamic model of a micro-grid including conventional and electronically interfaced distributed resources},
  author    = {Katiraei, Farid and Iravani, MR and Lehn, PW},
  journal   = {IET generation, transmission \& distribution},
  volume    = {1},
  number    = {3},
  pages     = {369--378},
  year      = {2007},
  publisher = {IET}
}

@article{iyer2010generalized,
  title     = {A generalized computational method to determine stability of a multi-inverter microgrid},
  author    = {Iyer, Shivkumar V and Belur, Madhu N and Chandorkar, Mukul C},
  journal   = {IEEE Transactions on Power Electronics},
  volume    = {25},
  number    = {9},
  pages     = {2420--2432},
  year      = {2010},
  publisher = {IEEE}
}

\end{document}